\documentclass[10pt]{article}

\usepackage{bm}
\usepackage[]{graphicx}
\usepackage{amsmath}
\usepackage{amssymb}
\usepackage{color}
\usepackage{soul}
\usepackage{nonfloat}
\usepackage[left=1.7in, top=1.7in]{geometry}

\begin{document}

\title{The \textquoteleft Swiss cheese\textquoteright \, instability of bacterial biofilms}

\author{Hongchul Jang, Roberto Rusconi, and Roman Stocker\\ \\Department of Civil \& Environmental Engineering, 
\\Massachusetts Institute of Technology \\ Cambridge, Massachusetts 02139, USA}

\maketitle
\thispagestyle{empty}
\pagestyle{empty}

\date{}

\vspace{-0.4in}

\begin{abstract}
We demonstrate a novel pattern that results in bacterial biofilms as a result of the competition between hydrodynamic forces and adhesion forces. After the passage of an air plug, the break-up of the residual thin liquid film scrapes and rearranges bacteria on the surface, such that a \textquoteleft Swiss cheese\textquoteright \, pattern of holes is left in the residual biofilm.
\end{abstract}
\setlength{\parskip}{12pt}

Bacteria often adhere to surfaces, where they develop polymer-encased communities (biofilms) that display dramatic resistance to antibiotic treatment. The permanence of biofilms on surfaces thus poses serious risks of infections and transmission of pathogens so that a better understanding of bacterial detachment from biofilms using shear or air bubbles may lead to novel strategies for biofilm disruption and removal.
Using micro-contact printing, we created hydrophobic patches on glass substrates, which strongly favor bacterial attachment and on which biofilms rapidly develop. After the establishment of a biofilm patch, we inject a controlled air plug and observed its effect on the biofilm patch.

After 8 h of biofilm growth, the passage of an air plug causes the formation of an intriguingly regular pattern where regions of surface-adhering bacteria alternate with \textquoteleft holes\textquoteright \, i.e. regions from which bacteria have been. This Swiss-cheese pattern results from a delicate balance between hydrodynamic forces and adhesion forces, as demonstrated by experiments with 4 h and 12 h old biofilm patches. Bacterial attachment after 4 h growth was weak, as demonstrated by the nearly complete detachment of the biofilm after the passage of an air plug. Bacterial attachment after 12 h, on the other hand, was so strong that an air plug produced almost no detachment. High-resolution videos of the dynamics of hole formation show that the residual liquid film left after passage of the air plug ruptures at discrete locations and the ensuing movement of the local contact line  scrapes bacteria outwards, creating holes.

This new pattern of biofilm rearrangement needs to be considered when shear-based removal strategies are designed, as biofilm removal might be confused with biofilm rearrangement. 
Furthermore, antibiotic treatment might be affected by this previously unrecognized spatial biofilm pattern in ways that will be interesting to characterize. 

\begin{figure}[below]
\begin{center}
\includegraphics[keepaspectratio,width=5.5 in]{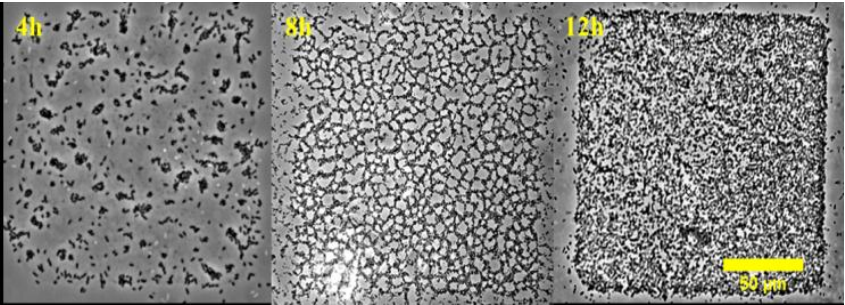} 
\end{center}
\label{Fig}
\end{figure}

\end{document}